 \documentclass[12pt]{amsart}

\usepackage{amsfonts,amssymb,amsmath} 
\usepackage{epsfig,graphicx}
\usepackage{mathptmx}      
\usepackage{mathtools}
\makeatletter

\renewcommand \thesection {\@arabic\c@section .}
\renewcommand\thesubsection   {\thesection \@arabic\c@subsection .}
\renewcommand\thesubsubsection{\thesubsection \@arabic\c@subsubsection .}
\renewcommand\theparagraph    {\thesubsubsection \@arabic\c@paragraph .}
\makeatother

\newcounter{subequation}
        {\addtocounter{equation}{-1}%
        \stepcounter{subequation}%
        \begin{equation}}%
        {\end{equation}%
}

\newcommand\abs[1]{{| #1 |}}

\newcommand{\rmd}{{\, \mathrm d}}
\newcommand{\rme}{{\, \mathrm e}}

\newcommand{\cH}{{\mathcal H}}

\newcommand{\cR}{{\mathcal R}}

\newcommand{\NN}{{\mathbb N}}

\newcommand{\RR}{{\mathbb R}}
\newcommand{\SSb}{{\mathbb S}}

\newcommand{\bfB}{{\mathbf B}}
\newcommand{\bfM}{{\mathbf M}}
\newcommand{\bfU}{{\mathbf U}}
\newcommand{\bfV}{{\mathbf V}}

\newcommand{\qvar}[1]{\langle#1\rangle}
\newcommand{\rmdS}{\circ \rmd}

\newcommand{\esq}{{\mathbb{E}}}
\newcommand{\ninf}{{N\rightarrow\infty}}
\newcommand{\dinf}{{D\rightarrow\infty}}

\newcommand{\scalprod}[2]{\langle #1, #2 \rangle}

\newtheorem{defn}{Definition}[section] 

\newtheorem{Prop}[defn]{\sc Proposition}

\newtheorem{Theo}[defn]{\sc Theorem}

\begin{document}

\title[Ornstein-Uhlenbeck limit]{Ornstein-Uhlenbeck limit for the velocity process of an $N$-particle system interacting stochastically}


\author{Bruno V. Ribeiro \and Yves Elskens}

 \address{B.V. Ribeiro : 
				Instituto de F\'isica, Universidade de Bras\'{\i}lia,
 				CP: 04455, 70919-970 - Bras\'{\i}lia - DF, Brazil \\
 				CAPES Foundation, Ministry of Education of Brazil,
 				Bras\'ilia - DF 70040-020, Brazil \\
 			{\tt{brunovr@fis.unb.br}} 
			}
 \address{B.V. Ribeiro \and Y. Elskens : 
		    		Equipe turbulence plasma, case 321, 
		      	PIIM, UMR 7345 CNRS,
		       	Aix-Marseille universit\'e,
		       	campus Saint-J\'er\^ome, FR-13397 Marseille cedex 13 \\
		      { \tt{yves.elskens@univ-amu.fr}}
}
\date{\today}
\maketitle

\begin{abstract}\noindent
     An $N$-particle system with stochastic interactions is considered. 
     Interactions are driven by a Brownian noise term 
     and total energy conservation is imposed. 
     The evolution of the system, in velocity space, is a diffusion 
     on a $(3N-1)$-dimensional sphere with radius fixed by the total energy. 
     In the $N\rightarrow\infty$ limit, a finite number of velocity components 
     are shown to evolve independently and according to an Ornstein-Uhlenbeck process.
  \\
\\
\textbf{Keywords} : {particle in a heat bath $\bullet$
  propagation of chaos $\bullet$
  Kac program $\bullet$
  stochastic interactions in many dimensions $\bullet$
  diffusion on a sphere $\bullet$
  Ornstein-Uhlenbeck process $\bullet$
  Kac systems}
  %
\\
\textbf{PACS} :  {
     05.40.-a $\bullet$
     05.45.-a $\bullet$
     05.10.Gg 
      }
   %
\\
 \textbf{MSC} :  { 
      34F05 $\bullet$		
      60H10 $\bullet$	
      82C05 $\bullet$	
      60J70  
      }
\end{abstract}



\vfill \hrule
\smallskip
\noindent \textit{To Etienne Pardoux on his sixty-fifth birthday}
\par \smallskip
\noindent  \textbf{preprint submitted for publication} \\
  \par
\noindent \copyright{2013} The authors. Reproduction of this article, in its entirety,
for noncommercial purposes is permitted.

\newpage

\section{Introduction}
 \label{secIntro}

The ensemble average of the stochastic evolution of a Brownian test particle immersed in a heat bath of constant temperature $\theta$ is given by the Fokker-Planck equation with prescribed coefficients of diffusion and linear friction \cite{UhOr30}. The solution to this equation is well known and gives the evolution of the ensemble's probability density function in velocity space.

Kiessling and Lancellotti \cite{KL06} have shown that, if the coefficients of this partial differential equation for the Ornstein-Uhlenbeck process are chosen to be constant functionals of the solution itself, the equation can be reinterpreted to describe the kinetic evolution of an isolated $N$-particle system with stochastic interactions satisfying total mass, energy and momentum conservation\footnote{Carlen and Gangbo \cite{CG04} studied a very similar kinetic equation, though using different techniques.}. In their work, these authors consider an infinite ensemble 
of independent and identically distributed (i.i.d.) random $N$-particle velocity vectors $\left\lbrace \mathbf{V}_\alpha \right\rbrace_{\alpha=1}^\infty$, each representing a possible microstate of the system in velocity space (in $\RR^{3 N}$), while particle positions are considered to be uniformly distributed over a periodic box. By a suitable transformation on the $\mathbf{V}_\alpha$'s, they show that the analysis of this $N$-particle system with energy and momentum conservation can be done in terms of an $N'$-particle system with only energy conservation and $N' = N - 1$.


Our purpose in this work is to analyse the evolution of the same particle system with stochastic interactions driven by this Brownian noise term, but from a many-body perspective, i.e.\  we look at a single realization of the system and study its evolution pathwise. According to the arguments of \cite{KL06}, it suffices to study the system with only the energy conservation requirement, and so we do here, thus we drop the prime in $N'$ and refer to an $N$-particle system, considering the ``velocities'' of this effective model as if they related to actual particles 
(in the spirit of Kac' one-dimensional historical model \cite{Kac56}). This model is also related to the class of Kac systems in the terminology of \cite{CCL03} (see App.~\ref{secAppKac}).

In a 3-dimensional setting,  energy conservation  bounds the evolution of the system, in velocity space, 
to a $(D-1)$-dimensional sphere with radius defined by initial energy (with $D = 3N$ being the number of degrees of freedom of the system). 
Therefore, the evolution of the system in velocity space can be modeled by the Brownian motion on the given sphere. 
Thus, we take advantage of the works of Stroock \cite{St71} and Brillinger \cite{Br97}, 
where the random evolution of a single particle on a sphere was modeled with stochastic integrals, 
to write the proper differential equations for the evolution of the $D$-dimensional vector $\mathbf{V}$ 
of all the velocity components for the $N$ particles. Singling out one component of $\mathbf{V}$ (call it $V_1$), 
we show it to converge, in the $\dinf$ limit, to an Ornstein-Uhlenbeck process \cite{UhOr30}, 
which is commonly used to describe the evolution of a single particle under ``white noise" and friction.

\section{Outline and strategy}
 \label{secOut}

We look at the diffusion of a $3$-dimensional $N$-particle system which obeys conservation of energy. This requirement bounds the evolution of the system, in velocity space, to a $(3N-1)$-dimensional sphere with radius defined by initial energy.
The dynamics of the system is given by a single stochastic differential equation for a ($D=3N$)-dimensional velocity vector 
($\mathbf{V}$) driven by a $D$-dimensional Brownian noise term introduced in Sec.~\ref{secDiff} 

A quick look at this evolution equation shows  that a single component of $\mathbf{V}$ evolves independently of the remaining \textit{directions} according to an Ornstein-Uhlenbeck process driven by a single noise term along the same direction, when this component is small enough.
Our focus, however, is on studying the limiting process for the components of $\mathbf{V}$ when these are of order of unity. Formally, we define a reference one-dimensional Ornstein-Uhlenbeck process, and show that the evolution of a single component of $\mathbf{V}$ converges (in probability) to this reference process as $\dinf$. 
This is properly stated in Sec.~\ref{secConv} 
It is proved in Sec.~\ref{secProof}

 This proof is easily extended to any finite  $d$-dimensional process $(d < D)$ constructed from $d$ components of the original diffusion. Thus, we show that these processes are independent and identically distributed (i.i.d.) in the $\dinf$ limit, thereby proving the propagation of initial independence of $d$ particles, viz.\ the ``propagation of molecular chaos'' or of ``Boltzmann's property'' (\cite{Kac56,Kac59}, see the recent review by Mischler and Mouhot \cite{MM11} for an extensive discussion), in Sec.~\ref{secnPart} 
 Vakeroudis and Yor's recent central limit theorem on paths \cite{VY11} may also provide further insight into the evolution of $d \gg 1$ degrees of freedom in the $D \to \infty$ limit.
 
The system with total energy and momentum conservation is considered in Sec.~\ref{secEnMo} We model its dynamics with a stochastic differential equation for the $3N$-dimensional velocity vectors driven by a $3N$-dimensional Brownian noise term. Here, we use the original $N$ particles (before dropping the prime) and show that the velocity processes of any finite number $m$ of particles, in the $\ninf$ limit, converge to $m$ independent three-dimensional Ornstein-Uhlenbeck processes.

In App.~\ref{secApp2rep},  we sketch the answer to a question raised by Kiessling and Lancellotti \cite{KL06} 
on the explicit characterization of the many-body stochastic process.
In App.~\ref{secAppKac}, we show how this model relates to the class of Kac systems.

\section{Diffusion on a sphere}
 \label{secDiff}

Consider a system of $N$ particles. The particles exchange momentum 
according to some interaction between them, 
and total energy is assumed to be conserved.

Let $D=3N$ denote the number of degrees of freedom of the system. 
The system state is described by the $D$-dimensional vector
\begin{equation}
	\mathbf{V} = (V_1,V_2,...,V_{D}),
\label{rvec}
\end{equation}
where the $V_i$'s are all the $D=3N$ components of the three-dimensional vectors 
$\bar{V}_k = (V_{3k-2}, V_{3k-1}, V_{3k})$ 
representing the velocity of the $k$-th particle in $\mathbb{R}^3$.
Conservation of energy requires the vector $\mathbf{V}$ to assume values 
in the $(3N-1)$-dimensional manifold of total energy $N e_0$
\begin{equation}
	\mathbb{M}^{3N-1}_{e_0} 
	= \left\lbrace (\bar{V}_1, \ldots \bar{V}_N) \ :\ \sum_{i=1}^{3N} \frac{|V_i|^2}{2} = N e_0 \right\rbrace ,
\label{capm}
\end{equation}
viz.\  the $(3N - 1)$-dimensional sphere with radius $\sqrt{2 N e_0}$.

Starting from an arbitrary initial data, the evolution of vector $\mathbf{V}$ 
is modeled by the Brownian motion on the sphere, given by the stochastic differential equation \cite{Br97}
\begin{equation}
	\rmd \mathbf{V} = \lambda \sigma(\mathbf{V})\cdot \rmdS \mathbf{B},
\label{strasp}
\end{equation}
where $\lambda$ is the noise amplitude, 
\begin{equation}
	\sigma_{ij}(\mathbf{V}) = \delta_{ij} - \frac{V_iV_j}{|\mathbf{V}|^2}
\label{sij}
\end{equation}
is the orthogonal projector onto the hyperplane orthogonal to $\mathbf{V}$,
\begin{equation}
	\mathbf{B} = (B_1,B_2,...,B_D)
\label{bmn}
\end{equation}
is the standard $D$-dimensional Brownian motion, and $\rmdS$ denotes the use of Stratonovich calculus \cite{Ok10}. Although the Stratonovich view-point seems physically natural in the sense that it leads to the same chain rules as classical calculus and occurs in taking limits from ordinary differential equations driven by smooth noises \cite{WoZa65}, we shall change view-points and prefer working with the It\^o representation. This representation has some advantages to our purposes, for instance, It\^o integrals are well known to be martingales \cite{Ok10}, 
which is crucial for stochastic analysis and computing expectations. The disadvantage is that It\^o integrals do not transform so nicely under changes of variables, e.g.\  the chain rule for It\^o differentials involves second order terms \cite{Ok10,KlPlSc03}.

Therefore, we work with the spherical Brownian motion as the solution of the It\^o differential equation \cite{St71,Pr05}
\begin{equation}
	\rmd\mathbf{V} 
	= \lambda \sigma(\mathbf{V})\cdot \rmd\mathbf{B} - \lambda^2 \frac{D-1}{2} b(\mathbf{V})\rmd t,
\label{itosp}
\end{equation}
where
\begin{equation}
	b(\mathbf{V}) = \frac{\mathbf{V}}{|\mathbf{V}|^2}.
\label{sb}
\end{equation}
The process (\ref{strasp}) remains on the surface of the sphere, 
the first (It\^o) term in the right hand side of (\ref{itosp}) displaces the velocity 
inside the tangent plane to the sphere, 
and the last term in (\ref{itosp}) may be thought of as pulling the process back onto the sphere \cite{Br97}. 
Introducing  $\theta := 2 e_0 / 3$, 
we rescale time to $t' = \lambda^2 t / \theta$, 
the Wiener process to $\mathbf{B}' = \lambda  \theta^{-1/2} \mathbf{B}$ 
and velocity to $\mathbf{V}' = \theta^{-1/2} \mathbf{V}$, so that, dropping primes, 
the system (\ref{itosp})-(\ref{sb}) reduces to 
\begin{equation}
	\rmd\mathbf{V} 
	= 
	\sigma(\mathbf{V})\cdot \rmd\mathbf{B} -  \frac{D-1}{2 D} \mathbf{V} \rmd t .
\label{itospred}
\end{equation}
Our probability space is the standard space of $D$-dimensional continuous brownian motions 
(see e.g.\ sec.\ I.3 in \cite{Pr05}).

\section{Convergence to the Ornstein-Uhlenbeck process}
 \label{secConv}

We single out one component of $\mathbf{V}$ to study its evolution. 
Evidently, this evolution will have a contribution from all the other directions 
(the $D-1$ remaining dimensions), so we define
\begin{equation}
	\mathbf{U} = \mathbf{V} - V_1 \hat{e}_1,
\label{fstset}
\end{equation}
where $\hat{e}_i$ is the unit vector along the direction of the $i$-th dimension. A simple rewriting of (\ref{itospred}) leads to the system
\begin{eqnarray}
	\rmd V_1 
	& = & 
	\sigma_1(V_1) \rmd B_1 
	- \frac{D-1}{2}b_1(V_1) \rmd t 
	- \mathbf{H}(\mathbf{U},V_1)\cdot \rmd \mathbf{B}_U  , 
  \label{ev1}
  \\	
     \rmd \mathbf{U} 
     & = & 
     \sigma_U(\mathbf{U}) \cdot \rmd \mathbf{B}_U 
     - \frac{D-1}{2}\mathbf{b}_U(\mathbf{U}) \rmd t 
     - \mathbf{H}(\mathbf{U},V_1)\rmd B_1 ,
  \label{evu}
\end{eqnarray}
given
\begin{eqnarray}
	\sigma_1(V_1) =& 1 - \frac{V_1^2}{|\mathbf{V}|^2},\ \ b_1(V_1) = \frac{V_1}{|\mathbf{V}|^2},\ \ \mathbf{H}(\mathbf{U},V_1) = \frac{V_1}{|\mathbf{V}|^2} \mathbf{U}   ,  \nonumber \\
	(\sigma_U(\mathbf{U}))_{ij} =& \delta_{ij} - \frac{U_iU_j}{|\mathbf{V}|^2}, \ \ \mathbf{b}_U(\mathbf{U}) = \frac{\mathbf{U}}{|\mathbf{V}|^2}, \ \ \mathbf{B}_U = (B_2,...,B_D)
\label{defs}
\end{eqnarray}
with the constraint $|\mathbf{V}| = \sqrt{D}$.

One obtains a simple estimate of the evolution of $V_1$ 
by first looking at the limit of small values for $V_1$ ($V_1 \sim 0$). 
In such a limit,
\begin{equation}
	\sigma_1(V_1) \sim 1 , \ \ \mathbf{H}(\mathbf{U},V_1) \sim 0.
\label{smallv1}
\end{equation}

So, the resulting evolution equation would be
\begin{equation}
	\rmd V_1 \simeq \rmd B_1 - \frac{D-1}{2|\mathbf{V}|^2}V_1 \rmd t,
\label{sv1}
\end{equation}
which 
reads, in the $\dinf$ limit, 
\begin{equation}
	\rmd V_1 \simeq \rmd B_1 - \frac{V_1}{2}\rmd t,
\label{sv1bn}
\end{equation}
which has as solution the standard Ornstein-Uhlenbeck process.


Let us now look at the general evolution of $V_1$ in the $\dinf$ limit, i.e.\  we want to know what is the limiting process defined by (\ref{ev1}) when $V_1$ is of order unity. From the result (\ref{sv1bn}), we can expect this process to, at least, be similar to the Ornstein-Uhlenbeck process. Our main result is expressed in the following

\begin{Theo}
\label{Thm1}
Let $V_1$ be a component of $\mathbf{V}$ defined by equation (\ref{itospred}). 
For $|\mathbf{V}|=\sqrt{D}$, the process $V_1$ with initial data $c_1$ obeying (\ref{ev1}) converges, 
in the $\dinf$ limit, to the $1$-dimensional Ornstein-Uhlenbeck process with the same initial data and driving noise $B_1$, 
where convergence is in the mean-square sense in $C([0,T], \mathbb{R})$ for any time $T > 0$.

%
\end{Theo}

  Of course, given $c_1 \in \RR$, the finite-$D$ model makes sense only for $D \geq c_1^2$.

\section{Proof of Theorem \ref{Thm1}}
 \label{secProof}

We start our analysis by defining the \textit{reference} (or target for the limit) 1-dimensional Ornstein-Uhlenbeck process
	\begin{equation}
		\rmd V_1' = \alpha \rmd B_1 - \beta V_1' \rmd t, \quad \alpha,\beta \in \mathbb{R}.
	\label{v1prime}
	\end{equation}
driven by the same (\textit{driving}) noise $B_1$ as in (\ref{ev1}). 
With this, we introduce
\begin{equation}
	v = V_1 - V_1' .
\label{sv}
\end{equation}

We  compute the expectation of the square of this new process to grasp the evolution of $V_1$ 
compared to the evolution of the reference process. 
So, we start by writing explicitly
\begin{equation}
	v(t) 
	= 
	v_0 + \int_0^t\left[ (\sigma_1(V_1) - \alpha)\rmd B_1 
	               - \left( \frac{D-1}{2}b_1(V_1)-\beta V_1' \right)\rmd t 
	               - \mathbf{H}(\mathbf{U},V_1)\cdot \rmd\mathbf{B}_U 
	               \right],
\label{expv}
\end{equation}
where $v_0=v(t=0)$.

We set the parameters $\alpha = 1$ and $\beta = 1/2$. Thus, the expectation reads
\begin{equation}
	\esq[|v(t)|^2] = \esq\left[ \left( v_0 - \int_0^t\left(\frac{v(s)}{2}- \frac{V_1(s)}{2D} \right) \rmd s - \int_0^t \frac{V_1^2}{D}\rmd B_1 - \int_0^t \frac{V_1}{D} \mathbf{U}\cdot \rmd \mathbf{B}_U \right)^2 \right],
\label{eu2}
\end{equation}
which is readily bounded as
\begin{equation}
	\frac{\esq[|v(t)|^2]}{3} \leq \esq[|v_0|^2] + \esq\left[\left( \int_0^t\left(\frac{v(s)}{2}- \frac{V_1(s)}{2D} \right) \rmd s \right)^2 \right] +  \esq\left[\left(\int_0^t \frac{V_1}{D} \mathbf{V}\cdot \rmd\mathbf{B} \right) ^2 \right]. 
\label{fstbound}
\end{equation}

Using the property $\rmd \qvar{ B_i, B_j }  = \delta_{ij}\rmd t$, 
where $\qvar{ \bullet , \bullet }$ denotes the cross-variation process \cite{Ok10}, 
and the fact that $\esq\left[\int_0^t f(s) \rmd B_i(s) \right] =0 $, for any (adapted) function of time $f$, 
the last term in the r.h.s.\ can be rewritten as
\begin{equation}
	\sum_{i=1}^D \esq\left[ \left(\int_0^t \frac{V_1}{D} V_i \rmd B_i \right)^2  \right] 
	= \sum_{i=1}^D \esq\left[ \int_0^t \frac{V_1^2}{D^2}V_i^2 \rmd s \right]
	= \esq\left[\int_0^t \frac{V_1^2}{D} \rmd s \right] ,
\label{fltv1}
\end{equation}
where we used It\^o isometry in the intermediate step and recalled our constraint, $|\mathbf{V}|=\sqrt{D}$, 
in the last step. 

Going back to (\ref{fstbound}), we further bound
\begin{eqnarray}
  &&
	\frac{\esq[|v(t)|^2]}{3} 
	\nonumber \\
	&\leq & \esq[|v_0|^2] 
	  + \frac{1}{2}\esq\left[\left( \int_0^t v(s) \rmd s \right)^2\right]  
	  + \frac{1}{2}\esq\left[\left(\int_0^t \frac{V_1(s)}{D} \rmd s \right)^2 \right] 
	  +  \esq\left[\int_0^t \frac{V_1^2}{D} \rmd s \right] \nonumber \\
	&\leq &  \esq[|v_0|^2] 
	  + \frac{t}{2}\int_0^t \esq\left[v(s)^2\right] \rmd s   
	  + \left( \frac{t}{2D^2} +  \frac{1}{D} \right) \int_0^t \esq\left[V_1^2(s)\right] \rmd s , 
	  \nonumber \\
	  &&
\label{sndbound}
\end{eqnarray}
where we used Schwartz inequality on the second and third terms in the r.h.s. 
To get a closed estimate, we must now control how fast $\esq\left[V_1^2(s)\right]$ grows in time. 
For this purpose, a standard It\^o calculation leads to
\begin{equation}
	\rmd V_1^2 = 2V_1\rmd V_1 + \left(1 -\frac{V_1^2}{D}\right) \rmd t,
\label{itov12}
\end{equation}
which gives 
\begin{equation}
	\rmd \esq[V_1^2] 
	= \left(1 - \frac{\esq[V_1^2]}{D}\right) \rmd t \leq \rmd t,
\label{esqitov1}
\end{equation}
so that
\begin{equation}
	\int_0^t \esq\left[V_1^2(s)\right] \rmd s 
	\leq 
	\int_0^t (c_1^2 + s) \rmd s 
	= 
	c_1^2 t + \frac{t^2}{2}.
\label{esqv1int}
\end{equation}

Thus, we  find a bound on $\esq[|v(t)|^2]$ independent of the component $V_1$,
\begin{eqnarray}
	\esq[|v(t)|^2] 
	& \leq &  
	3 \esq[|v_0|^2] + \frac{3t}{2} \int_0^t \esq\left[v(s)^2\right] \rmd s 
	  + 3  \left( \frac{t}{2D^2} +  \frac{1}{D} \right)  ( c_1^2 t + \frac{t^2}{2} )
	   \nonumber \\ &&
      \\
      & \leq & 
	3 \esq[|v_0|^2] + \frac{3}{2 D} (1 + \frac{T}{2D}) (2 c_1^2 T + T^2) 
	+ \frac{3 T}{2} \int_0^t \esq\left[v(s)^2\right] \rmd s 
\label{boundwt}
\end{eqnarray}
over the interval $[0,T]$ for any $T > 0$.
Gronwall's lemma \cite{Di80} then yields the bound 
\begin{equation}
	\esq[|v(t)|^2] 
	\leq 
	\left[3 \esq[|v_0|^2] + \frac{3}{2 D} (1 + \frac{T}{2D}) (2 c_1^2 T + T^2)\right] \rme^{3 T t / 2} .
\label{gronv1t}
\end{equation}

For any finite $t$, we may set $T=t$ in this estimate. 
Therefore, in the $\dinf$ limit, we have for finite times
\begin{equation}
	\lim_{D \to \infty}  \esq[|v(t)|^2] 
	\leq 3\esq[|v_0|^2]\rme^{3 t^2 / 2},
\label{gronv1t2}
\end{equation}
ensuring that, given initial conditions such that $v_0=0$, the process $V_1$ converges in mean square 
to the reference \emph{Ornstein-Uhlenbeck} process in one dimension over $[0,T] \ \forall \ T < \infty$, 
thus proving the theorem.
\hfill \qed

Note that the type of convergence obtained (mean-square convergence \cite{KlPlSc03}) implies convergence in probability in $C([0,T], \mathbb{R})$ for any time $T > 0$.

\section{Motion of $d$ particles}
 \label{secnPart}

This result can be generalized if we consider, instead of the single component $V_1$, 
a finite $d$-dimensional vector $\mathbf{U}_1$ made up of the first $d$ components of $\mathbf{V}$. 
Define
\begin{equation}
	\mathbf{U}_1 = \sum_{i=1}^d V_i \hat{e}_i ,  
	\quad
	\mathbf{U}_2 = \sum_{i=d+1}^D V_i \hat{e}_i ,
\label{u1u2}
\end{equation}
for a fixed $d$, and rewrite (\ref{itosp}) as
\begin{equation}
	\rmd \mathbf{U}_1 = \sigma_1(\mathbf{U}_1) \cdot \rmd \mathbf{B}_1 - \frac{D-1}{2}\mathbf{b}_1(\mathbf{U}_1) \rmd t - \mathbf{H}(\mathbf{U}_1,\mathbf{U}_2)\cdot \rmd \mathbf{B}_2 ,
\label{evu1}
\end{equation}\begin{equation}
	\rmd \mathbf{U}_2 = \sigma_2(\mathbf{U}_2) \cdot \rmd \mathbf{B}_2 - \frac{D-1}{2}\mathbf{b}_2(\mathbf{U}_2) \rmd t - \mathbf{H}^T(\mathbf{U}_1,\mathbf{U}_2)\cdot \rmd \mathbf{B}_1 ,
\label{evu2}
\end{equation}
where $^T$ denotes the transpose and
\begin{eqnarray}
	({\sigma_i})_{mn} &=& \delta_{mn} - \frac{({\bfU_i})_m ({\bfU_i})_n}{|\mathbf{V}|^2}; \quad \mathbf{b}_i(\mathbf{U}_i) = \frac{\mathbf{U}_i}{|\mathbf{V}|^2};\quad i=1,2  \\
	(\mathbf{H})_{mn} &=& \frac{({\bfU_1})_m ({\bfU_2})_n }{|\mathbf{V}|^2} \\
	\mathbf{B}_1 &=& (B_1,\ \ldots \ , B_d), \quad \mathbf{B}_2 = (B_{d+1},\ \ldots \ , B_D).
\label{12defs}
\end{eqnarray}

From this point of view, we derive the following
\begin{Theo}
 Let $\mathbf{U}_1$ be the $d$-dimensional component of $\mathbf{V}$
 evolving according to equation (\ref{evu1}). 
 For $|\mathbf{V}|=\sqrt{D}$, the process $\mathbf{U}_1$ with initial data $\mathbf{U}_1(0) = \mathbf{c}_1 \in \RR^d$ 
 converges, in the $\ninf$ limit, 
 to the $d$-dimensional Ornstein-Uhlenbeck process with initial data $ \mathbf{c}_1$ and driving noise $\bfB_1$, 
 where convergence is in the mean-square sense in $C([0,T], \mathbb{R}^d) $ for any time $T > 0$.
\end{Theo}

\noindent \textit{Sketch of proof} : The proof parallels the previous one, we need only define the reference $d$-dimensional Ornstein-Uhlenbeck process $\bfU_1'$. 
For this case, the form of the final bound on $\esq[|\mathbf{u}(t)|^2]$, with $\mathbf{u}(t) = \bfU_1(t) - \bfU'_1(t)$, is identical to (\ref{gronv1t}), replacing $v$ with $\mathbf{u}$ and $c_1$ with $\mathbf{c}_1$.

Note that $\abs{{\mathbf{c}}_1}^2$ is independent of $D$ (typically $O(d)$), which still leads to a vanishing contribution from the second term in the r.h.s.\ of the equation corresponding to (\ref{gronv1t}).
\hfill \qed

\section{The energy and momentum conserving model}
 \label{secEnMo}

In this section, we return from this model 
with $N' = N-1$ effective particles to the more physical $N$-particle model, 
with interactions conserving energy and momentum. 
In this setting, the vector of velocity components $\mathbf{V}$ assumes values in the manifold
\begin{equation}
    \mathbb{M}^{3N-4}_{\bar{u}_0, e_0} 
    = \left\lbrace (\bar{V}_1, \ldots \bar{V}_N) \ :\ 
             \sum_{i=1}^{3N} \frac{|V_i|^2}{2} = N e_0\ ; \ \sum_{k=1}^{N} \bar{V}_k = N\bar{u}_0 \right\rbrace ,
\label{capmm}
\end{equation}
where $\bar{u}_0$ is the momentum per particle and the vectors $\mathbf{V}$ and $\bar{V}_k$ are similar 
(with the original $N$) to those defined in Section \ref{secDiff} 
This manifold corresponds to the $(3N-4)$-dimensional sphere of radius $\sqrt{2N\varepsilon_0}$ 
centered at $(\bar{u}_0, ... , \bar{u}_0)$ and embedded in the constant momentum plane, 
with $\varepsilon_0 = e_0 - |\bar{u}_0|^2/2$.

Diffusion on $\mathbb{M}^{3N-4}_{\bar{u}_0, e_0}$ may be defined with the Stratonovich differential equation
\begin{equation}
    \rmd \mathbf{V} = \lambda P(\mathbf{S}) \cdot \rmdS \mathbf{B} ,
\label{Wev}
\end{equation}
where the $S_i$'s are the components of the three-dimensional vectors 
\begin{equation}
    \bar{S}_k = \bar{V}_k - \bar{u}_0 = (S_{3k-2}, S_{3k-1}, S_{3k})
\label{wcm}
\end{equation}
and $P$ is the orthogonal projector onto $\mathbb{M}^{3N-4}_{\bar{u}_0, e_0}$ given by
\begin{equation}
	P_{ij}(\mathbf{S}) 
	=  \sum_k \sigma_{ik}(\mathbf{S}) \left(\delta_{kj} - \sum_{\gamma=1}^3 \frac{E_k^\gamma E_j^\gamma}{N} \right)
	=  \sigma_{ij}(\mathbf{S}) - \sum_{\gamma=1}^3 \frac{E_i^\gamma E_j^\gamma}{N},
\label{MPro}
\end{equation}
with the $3N$-dimensional vectors $E^\gamma, \gamma = 1,2,3$, composed of all the components of $N$ unit vectors in $\mathbb{R}^3$ along the $\gamma$ direction, viz.
\begin{eqnarray}
	E^1 &=& (1,0,0,1,0,0,...,1,0,0) \nonumber \\
	E^2 &=& (0,1,0,0,1,0,...,0,1,0) \\
	E^3 &=& (0,0,1,0,0,1,...,0,0,1) \nonumber
\end{eqnarray} 
and $\mathbf{B}$ is now the standard $3N$-dimensional brownian motion. 

Kiessling and Lancellotti \cite{KL06} use  Gram-Schmidt orthonormalization, $\mathbf{s} = \cR \mathbf{V}$,
\begin{eqnarray}
    \bar{s}_n &=& \sqrt{\frac{N-n}{N-n+1}}\left[ \bar{V}_n - \frac{1}{N-n}\sum_{i=n+1}^N \bar{V}_i \right]
    \quad \textrm{ for } 1 \leq n \leq N-1 , 
    \label{klmap} \\
    \bar{s}_N &=& \frac{1}{\sqrt{N}}\sum_{i=1}^N \bar{V}_i,
    \label{kln}
\end{eqnarray}
to map $\mathbb{M}^{3N-4}_{\bar{u}_0, e_0}$ to
\begin{equation}
    \left\lbrace (\bar{s}_1, \ldots \bar{s}_N) \ : \ \bar{s}_N = \sqrt{N}\bar{u}_0, \
    \sum_{i=1}^{N-1} \frac{|\bar{s}_i|^2}{2} = N\varepsilon_0 \right\rbrace .
\label{newsp}
\end{equation}
As the matrix $\cR$ is orthogonal and $\mathbf{B}$ is isotropic, 
$\cR \mathbf{B}$ is also a standard brownian motion in $\RR^{3N}$.
In this new representation, $P$ is simply the identity on the $3(N-1)$ first components, 
and zero on the last three ones. 

So, we can analyse the original $N$-particle system with energy and momentum conservation, 
in terms of the truncated $(\bar{s}_1, ..., \bar{s}_{N-1})$ vector, 
with only ``energy in the center of mass frame'' ($\varepsilon_0$) conservation, 
and the constant vector $(\bar{s}_N)$. 
In particular, for particle $n=1$, one finds \\
$ \bar V_1 = \bar{s}_1 \sqrt{(N-1)(N-2)}/N  + \bar{s}_N / \sqrt{N} $,
so that  $\bar V_1 - \bar{u}_0$ also tends to an Ornstein-Uhlenbeck process for $N \to \infty$.

Actually, one can also directly prove
\begin{Prop}
\label{Prop1}
 Consider $m$ particles in the $N$-body model (\ref{Wev})
 with given total momentum per particle $\bar{u}_0$ and energy in the center of mass frame $\varepsilon_0$.
 Let the particles initial velocities be $\bar{u}_0 + \bar{c}_k$ ($1 \leq k \leq m$).
 In the $\ninf$ limit, 
 their velocity processes $\bar{S}_k = \bar V_k - \bar{u}_0$ converge
 to $m$ independent three-dimensional Ornstein-Uhlenbeck processes, 
 with initial data $ \bar{c}_k$ and driving noise $\tilde{B}_k = (B_{3k-2}, B_{3k-1}, B_{3k})$,
 where convergence is in the mean-square sense in $C([0,T], \mathbb{R}^{3m}) $ for any time $T > 0$.
\end{Prop}

The proof follows closely that of Sec. \ref{secProof} To complete the proof, one must notice that $\sum_{k=1}^N \bar{S}_k = 0$ and, in the $\ninf$ limit, that the process
\begin{equation}
	\tilde{B}_1 
	- \frac{1}{N} \sum_{k=1}^N \tilde{B}_k
\end{equation}
converges, in the mean-square sense, to $\tilde{B}_1$. To show this, it suffices to use \eqref{fltv1} with the changes $V_1 V_i \rightarrow 1$ and $D \rightarrow N$.

\section{Conclusions}
 \label{secConc}

A $3$-dimensional $N$-particle system, in which particles interact stochastically due to a driving Brownian noise term, 
satisfying total energy ($N e_0$) conservation is modeled by a diffusion process 
on the sphere $\mathbb{M}^{3N-1}_{e_0}$. In the $\ninf$ limit, 
a single velocity component evolves independently of all the other velocity components for finite times. 
Furthermore, this ``tagged'' component converges to a $1$-dimensional Ornstein-Uhlenbeck process 
driven by the single noise component coupled directly to it. 
The role of friction in the usual Ornstein-Uhlenbeck model to keep the velocity component bounded 
is here played by total energy conservation and by the curvature of the sphere 
(equatorial bands have a larger area than polar caps).

Technically, convergence to the Ornstein-Uhlenbeck process is proven using standard arguments 
(see section 5.2 of \cite{Ok10}). In the language of \cite{KlPlSc03}, 
we obtain \textit{mean-square convergence} for the velocity process in (\ref{gronv1t2}). 
Recall that mean-square convergence implies convergence in probability.

An immediate extension is given for the $d$-dimensional velocity process. In the $\dinf$ limit, 
this is also shown to converge to the $d$-dimensional Ornstein-Uhlenbeck process. 
If at initial time the $d$ tagged components have independent data, 
the subsequent evolution of these $d$ components preserves their independence, 
which amounts to propagating molecular chaos.

For the total energy and momentum conserving model, 
we show that the velocities of $m$ tagged particles converge,
in the $\ninf$ limit, to $m$ independent three-dimensional Ornstein-Uhlenbeck process driven
by their respective independent three-dimensional noises. 

We leave to the reader or for a forthcoming work the discussion of the three-dimensional model of App.~\ref{secApp2rep}, 
where one will eliminate the self-coupling gyroscopic terms in $\Omega$ 
and show that, in the limit $N \to \infty$, 
one recovers the same limit. 
Such a model is considered, from the Fokker-Planck viewpoint, in section 4 of \cite{CCL03}.

\section*{Acknowledgements}
 \label{secAckn}
 
BVR thanks the \textit{Coordena\c{c}\~ao de Aperfei\c{c}oamento de Pessoal de N\'ivel Superior} (CAPES) for financing his stay at Aix-Marseilles University through the \textit{Programa Institucional de Bolsas de Doutorado Sandu\'iche no Exterior} (PDSE), process number 8510/11-3. YE thanks Michael Kiessling for drawing his attention to this problem. 
We thank Wendell Horton for fruitful discussions.
We thank the referees for useful comments and for drawing our attention to \cite{CG04}.

 \renewcommand{\thesection}{\Alph{section}}
 \setcounter{section}{0}
\section{Appendix : Two representations of the diffusion on the sphere, and a binary interaction interpretation}
 \label{secApp2rep}
 
Our representation of the diffusion on the sphere, with the It\^o differential equation (\ref{itospred}), 
describes the evolution of the velocity vector in terms of $D$ independent brownian driving noises $B_j$, $1 \leq j \leq D$,
each directly associated with one component of $\bfV$.
Kiessling and Lancellotti \cite{KL06} recall the representation of the Laplace-Beltrami operator on the unit sphere
$\SSb^{D-1} \subset \RR^D$ in the form
\begin{equation}
 \triangle_{\SSb^{D-1}} 
  = 
  \sum_{1 \leq k < l \leq D} (v_k \partial_{v_l} - v_l \partial_{v_k} )^2
  \label{lapsp}
\end{equation}
and  interpret the Fokker-Planck equation in terms of a model for their original $N = 1+ D/3$ particles 
with two-body and one-body interactions preserving momentum and energy. 
Here we translate this interpretation to a simple stochastic process, limiting ourselves to the $D/3$ 
 ``effective particles'' with interactions preserving only energy.

Diffusion on the sphere may be viewed as a succession of independent infinitesimal rotations. 
So, consider a family of $D(D-1)/2$ independent processes 
$\Omega_{kl}(.), 1 \leq k < l \leq D$, taken to be martingales with initial value $0$ and 
cross-variation  $\qvar{\Omega_{ij} , \Omega_{kl}} (t) = \delta_{ik}Ê\delta_{jl} \  t / D$. 
Complement $\Omega$ to an antisymmetric matrix, with 
$\Omega_{ij} = - \Omega_{ji}$, so that the differentials $\rmd \Omega_{ij}$ 
act as infinitesimal rotation generators, 
and consider the process $\bfV'$ defined by an initial data $\bfV'(0)$ with $\abs{\bfV'(0)} = \sqrt{D}$ 
and the Stratonovich differential equation
\begin{equation}
  \rmd V'_i = \sum_j V'_j \rmdS \Omega_{ij}  .
\label{eqAppStrato}
\end{equation}

Its Fokker-Planck operator, acting on measures $f$ 
on the sphere $\SSb^{D-1}_{\sqrt{D}}$ with radius $\sqrt{D}$, 
is the sum $L = - \frac{1}{2} \sum_{1 \leq i < j \leq D} R_{ij}^* R_{ij}$, 
where $R_{ij}$ is the vector field associated with $\Omega_{ij}$, and $R_{ij}^*$ is its adjoint operator on $\RR^D$. 
By (\ref{eqAppStrato}), the vector field is $R_{ij} = D^{-1/2} (v'_j \partial_{v'_i} - v'_i \partial_{v'_j})$. Then 
$R_{ij}^* = - D^{-1/2} ( \partial_{v'_i} (v'_j \ \cdot) - \partial_{v'_j} (v'_i \ \cdot) ) = - R_{ij}$, so that $L = \frac{1}{2D} \triangle_{\SSb^{D-1}}$.
As the generator determines the law of a diffusion, 
this proves that the process defined by (\ref{eqAppStrato}) is the same as the one defined in Sec.\ \ref{secDiff}

Now, comparing directly the differential equations is also interesting. So, from the Stratonovich form (\ref{eqAppStrato}), 
we deduce the It\^o equation for $\bfV'$,
\begin{eqnarray}
  \rmd V'_i 
  & = &
  \sum_j V'_j \rmd \Omega_{ij} 
  + \frac{1}{2} \sum_j \rmd \qvar{ V'_j , \Omega_{ij} }
  \label{eqAppIto1} \\
  & = & 
  \sum_j V'_j \rmd \Omega_{ij} 
  + \frac{1}{2} \sum_j \sum_{k \neq j} V'_k \rmd \qvar{ \Omega_{jk} , \Omega_{ij} }
  \nonumber \\
  & = & 
  \sum_j V'_j \rmd \Omega_{ij} 
  + \frac{1}{2} \sum_j \sum_{k \neq j} V'_k (-D^{-1} \delta_{jj}Ê\delta_{ik}) \rmd t 
  \nonumber \\
  & = & 
  \sum_j V'_j \rmd \Omega_{ij} 
  - \frac{1}{2} V'_i \,  (1 - D^{-1} ) \rmd t     .
  \label{eqAppIto}
\end{eqnarray}  
For the first equality, we used the relation between It\^o and Stratonovich integrals (see e.g.\ Sec.~V.5 in \cite{Pr05}) ; 
then we substitute the first term of the r.h.s.\ of (\ref{eqAppIto1}) into the cross-variation term, 
and finally we use the explicit cross-variation of $\Omega$.

In the final expression (\ref{eqAppIto}), the drift is exactly the one in (\ref{itospred}). 
The first term appears as coupling the $i$-th component of vector $\bfV'$ with all other components, 
and it is linear with respect to $\bfV'$ while (\ref{itospred}) involves a projection matrix $\sigma$ 
orthogonal to $\bfV$. 
As this first term is the differential of a martingale, say $\rmd M'_i = \sum_j V'_j \rmd \Omega_{ij} $, 
we now show that it is equivalent (in law) to the first term in (\ref{itospred}). 
Indeed, this martingale $\bfM'$ is further characterized by its cross-variation process, 
which has the differential
\begin{eqnarray}
  \rmd \qvar{ M'_i, M'_j} 
  & = &
  \sum_k \sum_l V'_k V'_l \rmd \qvar{\Omega_{ik} , \Omega_{jl} }
  \nonumber \\
  & = &
  \sum_k \sum_l V'_k V'_l D^{-1} (\delta_{ij} \delta_{kl} - \delta_{il}Ê\delta_{jk}) \rmd t
  \nonumber \\
  & = &
  D^{-1} (\delta_{ij} \sum_k {V'_k}^2 - V'_i V'_j) \rmd t
\end{eqnarray}
where one readily recognizes the projector entry $\sigma_{ij}(\bfV')$. 

For comparison, the first term in  (\ref{itospred}) defines a martingale $\bfM$ 
with $\rmd M_i = \sum_j \sigma_{ij}(\bfV) \rmd B_j$, 
so that its cross-variation satisfies
\begin{equation}
  \rmd \qvar{M_i, M_j} 
  =
  \sum_k \sum_l \sigma_{ik} \sigma_{jl} \rmd \qvar{B_k , B_l}
  =
  \sigma_{ij}(\bfV)\, \rmd t
\end{equation}
as follows from the cross-variation of $\bfB$ and the fact that $\sigma^2 = \sigma$. 

As a result, $\bfM'$ and $\bfM$ have the same cross-variation when generated from the same trajectory 
(assuming $\bfM(0) = \bfM'(0) = \mathbf{0}$ for definiteness), 
as befits processes $\bfV'$ and $\bfV$ having the same law 
(incidentally, these calculations  reformulate the fact that the Fokker-Planck generators are equal).
However, the brownian drivers underlying both processes differ, 
as those for $\bfV$ are defined from translations along the $D$ components directions 
while those for $\bfV'$ are associated with the rotation matrices acting on $\RR^D$. 
Interestingly, the non-abelian nature of rotation compositions is irrelevant to our discussion.

Kiessling and Lancellotti \cite{KL06} interpret their representation (A.5) of the Laplace-Beltrami operator 
on the sphere in terms of two types of contributions. Here, one associates with three components, 
$3k - 2 \leq i \leq 3k$, the cartesian components of the velocity $\bar V'_k$ of particle $k$ in $\RR^3$, 
so that  in the $\Omega_{ij}$'s one distinguishes the terms associated with different particles 
($\lceil i/3 \rceil \neq \lceil j/3 \rceil$ for the ceiling function $\lceil \cdot \rceil$), 
and terms coupling two velocity components of the same particle ($\lceil i/3 \rceil = \lceil j/3 \rceil$) 
so that the particle velocity vector just rotates in $\RR^3$ as under a gyroscopic force.
As they observe, the gyroscopic contribution to each particle velocity 
drops out in the $N \to \infty$ limit.

\section{Appendix : The Kac system property}
 \label{secAppKac}

Here we write $N$ for $D$ to follow the notations of Carlen, Carvalho and Loss closely.
Kac' model is a random walk on the sphere $\SSb^{N-1}$, 
where, at (Poisson distributed) random times, 
two particles are picked up at random, and their velocities $(V_i, V_j)$ 
suffer a random rotation by an angle $\theta$, according to a probability measure $\varrho(\theta) \rmd \theta$ 
in the $(\hat e_i, \hat e_j)$ plane, with $\varrho$ being continuous and even. 
Given a $\varrho_\varepsilon$ with support $[-\varepsilon \pi, \varepsilon \pi]$, 
this system can be modelled pathwise by a jump process for the velocity components $V_i$, 
driven by piecewise constant noise processes $\omega_{ij,\varepsilon} = - \omega_{ji,\varepsilon}$, 
with increments
\begin{equation}
	\Delta V_i = 
	\sum_{j \neq i} [ V_i \, (\cos  \Delta \omega_{ij, \varepsilon} - 1) 
	                          + V_j \, \sin  \Delta \omega_{ij, \varepsilon} ] 
\label{JumpProc}
\end{equation}
where $V(t)$ in the right hand side is understood as the left limit, $V(t-)$.
In the noise $\omega_{ij,\varepsilon} (t) = \sum_{k=1}^{K_{ij}(t)} \theta_{ij,k}$, 
the \textit{jumps} $\Delta \omega_{ij, \varepsilon} = \theta_{ij,k} = - \theta_{ji,k}$ 
are independent, distributed with law $\varrho_\varepsilon$, 
and their number $K_{ij}(t)$ in $[0,t[$ is Poisson distributed 
with rate $\lambda_\varepsilon = 1/\varepsilon^2$. 
In the limit $\varepsilon \to 0$, one can show that the noise becomes Brownian, 
and solutions to equation \eqref{JumpProc} converge to those of  \eqref{eqAppStrato}. 
%

In the following, we show directly that our model has all features of Kac systems, 
which are defined in \cite{CCL03} as systems of probability spaces, 
depending on $N \in \NN_0$, verifying four features. 
The first feature is the invariance, under particle labeling permutations, 
of the stationary measure $\mu_N$ 
which is the microcanonical measure on the sphere $X_N = \SSb^{N-1}_{\sqrt{N}}$. 
The second feature characterizes $\nu_N$ as the marginal distribution of the component $j$ of interest 
obtained by the projection $\pi_j$, in our case $\pi_j(\mathbf{V}) = V_j \in Y_N = [-\sqrt{N}, \sqrt{N}]$ : 
\begin{equation}
  \rmd \nu_N (y) 
  = \frac{\abs{\SSb^{N-2}}}{\abs{\SSb^{N-1}}} \left(1 - \frac{y^2}N \right)^{(N-3)/2} N^{-1/2} \rmd y
  \label{nuN}
\end{equation}
where ${\abs{\SSb^{N-1}}}$ is the measure of the unit $(N-1)$-dimensional sphere.
The third feature introduces the lift from $N-1$ components in $X_{N-1}$ 
along with one additional component in $Y_N$, 
in a form consistent with the definition of the projection $\pi_j$, viz.\ for $j=N$
\begin{equation}
  \phi_N (u, y) = ( s_N(y) u , y) 
  \label{phiN}
\end{equation}
with the scaling function $s_N(y) := \sqrt{\frac{N-y^2}{N-1}}$ projecting $u$ onto the sphere 
with radius $\sqrt{N-y^2}$, so that $(\mu_{N-1} \otimes \nu_N)(\phi_j^{-1}(A)) = \mu_N(A)$ 
for any measurable $A \subset X_N$.

The fourth feature deals with the Markov transition operator $Q_N$, 
which occurs in the generator of the evolution for the probability distribution function, 
such that $\partial_t f = (1/\tau_N)(Q_N - I_N) f$ where $I_N$ is the identity 
and $\tau_N$ is a characteristic time, e.g.\ $\tau_N = 1/N$ for Kac' original model 
(see section 1 in \cite{CCL03}). 
In our case, the Fokker-Planck equation yields
\begin{equation}
  Q_N = I_N + \frac{1}{2 N} \sum_{k=1}^N \sum_{i=1}^N  \partial_{v_i} \sigma_{ik}(v) \partial_{v_k} 
\label{defQ}
\end{equation}
from the Stratonovich formulation, recalling that $\sigma$ is symmetric and idempotent.

Then we must check that for any $f \in \cH_N := L^2(X_N,  \mu_N)$ 
\begin{equation}
  \scalprod{ f }{ Q_N f }_{\cH_N} 
  = \frac{1}{N} \sum_{j=1}^N \int_{Y_N} \scalprod{ f_{j,y} }{ Q_{N-1} f_{j,y} }_{\cH_{N-1}} \rmd \nu_N(y)
  \label{feature4}
\end{equation}
where $f_{j,y} (u) := f(\phi_j(u,y))$ for $y \in Y_N$ and $u \in X_{N-1}$ ; 
in our case, it simply reads $f_{N,y} (u) = f( s_N(y) u_1, \ldots s_N(y) u_{N-1}, y)$, 
and indices $j<N$ select similarly the $j$-th component instead of the $N$-th one. 

With our $Q$, $f$ must be twice differentiable, so we take $f \in H^2(X_N, \mu_N)$. 
Now, condition (\ref{feature4}) is linear in $Q$, so it suffices to prove it for $I$ and $(Q-I)/\tau$ separately. 
First, for $I$, note that
\begin{eqnarray}
  \int_{Y_N} \scalprod{ f_{j,y} }{ f_{j,y} }_{\cH_{N-1}} \rmd \nu_N(y)
  & = & \int_{Y_N} \left[ \int_{X_{N-1}} \abs{f(\phi_j(u,y))}^2 \rmd \mu_{N-1}(u)\right]  \rmd \nu_N(y)
  \nonumber  \\
  & = & \scalprod{ f }{ f }_{\cH_N} 
\end{eqnarray}
where the first equality follows from the definition of the Hilbert scalar product, and the second 
from Fubini and feature 3. 
Then summing for $1 \leq j \leq N$ and dividing by $N$ shows (unsurprisingly) that $I$ fulfills feature 4.


For $Q-I$, we use the representation \cite{KL06} of the Laplace-Beltrami operator on the sphere $X_N = \SSb^{N-1}_{\sqrt{N}}$
\begin{equation}
  2  (Q_N - I_N) / \tau_N 
  =   (1/N) \sum_{k<l}^N (v_k \partial_{v_l} - v_l  \partial_{v_k})^2 \, .
\label{lapspN}
\end{equation}
Then,
\begin{eqnarray}
  && \scalprod{ f_{j,y} }{ 2 (Q_{N-1} - I_{N-1}) f_{j,y} }_{\cH_{N-1}} 
  \nonumber \\
  & = &  \frac{\tau_{N-1}}{N-1}  \sum_{\substack{k < l\\ k \neq j,\  l \neq j}} \int_{X_{N-1}} 
      ( f_{j,y}(u) (u_k \partial_{u_l} - u_l  \partial_{u_k})^2 f_{j,y}(u) ) \rmd \mu_{N-1} (u),
\end{eqnarray}
where we note that the operator $u_k \partial_{u_l} - u_l \partial_{u_k}$ is homogeneous, so that the factor $s(y)$ in the change of variable (\ref{phiN}) from $u$ to $v = \phi_j(u,y)$ will be absorbed. Thus,
\begin{eqnarray}
  &&
  \sum_{j=1}^N \int_{Y_N} \scalprod{ f_{j,y} }{ 2 (Q_{N-1} - I_{N-1}) f_{j,y} }_{\cH_{N-1}} \rmd \nu_N(y)
  \nonumber \\
  & = & 
  \sum_{j=1}^N \frac{\tau_{N-1}}{N-1}  \sum_{\substack{k < l\\ k \neq j,\  l \neq j}} 
    \int_{Y_N} \int_{X_{N-1}} 
      ( f_{j,y}(u) (u_k \partial_{u_l} - u_l \partial_{u_k})^2 f_{j,y}(u) ) \rmd \mu_{N-1} (u) \rmd \nu_N(y)
  \nonumber \\
  & = & 
  \sum_{j=1}^N \frac{\tau_{N-1}}{N-1}  \sum_{\substack{k < l\\ k \neq j,\  l \neq j}} 
    \int_{X_N} 
      ( f(v) (v_k \partial_{v_l} - v_l \partial_{v_k})^2 f(v) ) \rmd \mu_{N} (v),
  \nonumber \\
  &  & 
\end{eqnarray}
where the second equality holds from Fubini and feature 3.
Note that we may interchange the sums and keep the restrictions on indices by changing
\begin{equation}
	\sum_{j=1}^N \sum_{\substack{k < l\\ k \neq j,\  l \neq j}}^N 
	= \sum_{k<l}^N \sum_{\substack{j = 1\\ j \neq k,l}}^N .
\label{PerSum}
\end{equation}
Thus, we have
\begin{eqnarray}
  &&
  \frac{1}{N}\sum_{j=1}^N \int_{Y_N} \scalprod{ f_{j,y} }{ 2 (Q_{N-1} - I_{N-1}) f_{j,y} }_{\cH_{N-1}} \rmd \nu_N(y)
  \nonumber \\
  & = &  
  \sum_{k<l}^N \sum_{\substack{j = 1\\ j \neq k,l}}^N \frac{\tau_{N-1}}{N-1}  \left[ \frac{1}{N} 
    \int_{X_N} 
      ( f(v) (v_k \partial_{v_l} - v_l \partial_{v_k})^2 f(v) ) \rmd \mu_{N} (v) \right] .
\end{eqnarray}

Now, the summand depends only on $k$ and $l$, so that the $N-2$ values of $j$ contribute with the same result. Furthermore, we recognize the resulting $(k,l)$ sum  as $\scalprod{ f }{ 2 \tau_N^{-1} (Q_N - I_N) f}_{\cH_{N}}$.
Hence,
\begin{equation}
	\frac{1}{N}\sum_{j=1}^N \int_{Y_N} \scalprod{ f_{j,y} }{ 2 (Q_{N-1} - I_{N-1}) f_{j,y} }_{\cH_{N-1}} \rmd \nu_N(y) 
	=
  \frac{(N-2) \tau_{N-1}}{(N-1) \tau_N} \scalprod{ f }{ 2 (Q_N - I_N) f}_{\cH_{N}} .
\end{equation}


To meet feature 4, it suffices to set $\tau_N = 1/(N-1)$, which scale as O($1/N$) as $N \to \infty$.
Note that $1/(N \tau_N)$ also describes the $N(=D)$-dependence of the drift term in the It\^o equation (\ref{itospred}).

In our pathwise formulation, 
feature 4 relates the stochastic differential equation (\ref{eqAppStrato}) of the $N$-particle system
to the equations for $(N-1)$-particle subsystems as follows. 
First, the $(N-1)$-particle subsystem, with particle $j$ removed, 
obeys the equation for $u_{i , \not j} = V_i / s_N(V_j)$ ($i \neq j$) 
\begin{equation}
  \rmd u_{i , \not j} = \tau_{N-1}^{1/2} \sum_k u_{k , \not j} \rmdS \Omega_{ik, \not j}  
\label{eq_u_Strato}
\end{equation}
where the $\Omega_{ik, \not j}$'s  are independent brownian motions with variance $t/(N-1)$ 
(see App.~\ref{secApp2rep} for the denominator $N-1$, for the diffusion on $X_{N-1}$) for 
$1 \leq i < k \leq N, i \neq j, k \neq j$, and $\Omega_{ki, \not j} = - \Omega_{ik, \not j}$. 
Then the generator on the right hand side of (\ref{feature4}) corresponds formally to the evolution equation
\begin{eqnarray}
  \rmd V_i 
  & = & 
  \frac{1}{\sqrt{N}}Ê\sum_{j=1}^N \sum_{l \neq j} 
      \left(\frac{\partial V_i}{\partial u_{l , \not j}}\right)_{V_j} \rmdS u_{l , \not j} 
  \\ & = &
  \frac{1}{\sqrt{N}}Ê\sum_{j=1}^N (1 - \delta_{ij}) \, s_N(V_j) \rmdS u_{i , \not j} 
  \\ & = &
  \frac{\tau_{N-1}^{1/2} }{\sqrt{N}}Ê\sum_{j \neq i } s_N(V_j)  \sum_{k \neq j} u_{k , \not j} \rmdS \Omega_{ik, \not j}  
  \\ & = &
  \frac{\tau_{N-1}^{1/2} }{\sqrt{N}}Ê\sum_{k \neq i} \sum_{j \neq i, j \neq k}  V_k \rmdS \Omega_{ik, \not j}  \, ,
\label{eq_feat4_Strato}
\end{eqnarray}
where the first equality is our interpretation of (\ref{feature4}), 
the second follows from the projection of $X_N$ onto $X_{N-1}$ and the lift (\ref{phiN}), 
the third follows from (\ref{eq_u_Strato}) and the fourth again from the lift (\ref{phiN}) 
(and $k \neq i$ because $\Omega$ is antisymmetric). 
To obtain (\ref{eqAppStrato}) formally, it suffices to define 
$\Omega_{ik} := (\tau_{N-1} / \tau_N)^{1/2} N^{-1/2} \sum_{j \neq i, j \neq k} \Omega_{ik, \not j}$~; 
for this $\Omega_{ik}$ to be the brownian motion with variance $t/N$, we set again $\tau_N = 1/(N-1)$ so that 
$\tau_{N-1} /  \tau_N = (N-1)/(N-2)$ since the sum over $j$ has only $N-2$ terms. 

Note that this latter check for feature 4 works pathwise, with an explicit construction 
of the new driving noises $\Omega$ for $N$ degrees of freedom from the subsystems' driving noises. 
In contrast, working with the generators $(Q-I)/N$ ensures only equivalence in law 
of the processes constructed from the subsystems with the $N$-particle process.




%
\end{document}